\documentclass{iopart}
  \usepackage{color}
  \usepackage{newlfont}
  \usepackage{graphicx}
  \usepackage{amssymb}
  \expandafter\let\csname equation*\endcsname\relax
  \expandafter\let\csname endequation*\endcsname\relax
  \usepackage{amsmath}
  \usepackage[latin1]{inputenc}

\newcommand{\ket}[1]{\mbox{$ | #1 \rangle $}}
\newcommand{\bra}[1]{\mbox{$ \langle #1 | $}}

\newcommand{\MD}{\textrm{\tiny MD}}

\newcommand{\be}{\begin{eqnarray}}
\newcommand{\ee}{\end{eqnarray}}
\newcommand{\cm}{\text{\em\tiny CM}}

\begin{document}

\title{Kinematics and dynamics in noninertial quantum frames of reference}
\author{R. M. Angelo}
\address{Departamento de Física, Universidade Federal do Paraná, P.O. Box 19044, 81531-980, Curitiba, PR, Brazil.}
\author{A. D. Ribeiro}
\address{Departamento de Física, Universidade Federal do Paraná, P.O. Box 19044, 81531-980, Curitiba, PR, Brazil.}

\begin{abstract}
From the principle that there is no absolute description of a physical state, we advance the approach according to which one should be able to describe the physics from the perspective of a quantum particle. The kinematics seen from this frame of reference is shown to be rather unconventional. In particular, we discuss several subtleties emerging in the relative formulation of central notions such as vector states, the classical limit, entanglement, uncertainty relations, and the complementary principle. A Hamiltonian formulation is derived as well which correctly encapsulates effects of fictitious forces associated with the accelerated motion of the frame. Our approach shows, therefore, how to formulate nonrelativistic quantum mechanics within noninertial reference frames which can be consistently described by the theory, with no appeal to classical elements.
\end{abstract}


\maketitle

\section{Introduction}

It is an amazing physical phenomenon that a particle can be {\em delocalized} by a slit. In fact, in diffraction experiments the slit prepares a particle in a state of spatial superposition relative to the laboratory.  The slit and the detectors, on the other hand, have well-defined positions and velocities relative to the laboratory, as they are rigidly attached to it. But are these quantum notions of localization {\em absolute}? The fact that position is a {\em relational} physical quantity impels us to answer this question in the negative. Indeed, if the particle was granted the right to be a frame of reference, then the interpretation would be rather different: it is the particle itself that {\em prepares} the slit in a delocalized state. That is, from the perspective of the particle---in this capacity, a {\em quantum reference frame}---the slit, the detectors, and even the experimentalist, are not spatially localized.

The notion of quantum frame of reference has shown to be crucial in yielding a deeper understanding of important problems in quantum physics. Some examples are representative. In 1957, starting from considerations about the physics seen by two different observers, one of them being part of the physical system described by the other, H. Everett stated his {\em relative state} formulation of quantum mechanics~\cite{everett57}. A decade later, Aharonov and Susskind~\cite{aharonov155,aharonov158} established a relation between quantum frames of reference and superselection rules associated with both angular momentum and charge. Later on, Aharonov and co-authors elaborated on the notions of vector potentials and inertial forces in moving frames~\cite{aharonov73,aharonov74} and presented, in Ref.~\cite{aharonov84}, a Hamiltonian formulation for the physics seen from the perspective of a quantum particle. More recently, many works have investigated fundamental properties of quantum reference frames \cite{pullin04,poulin06,poulin07,mackenzie12,BSLB08,GP08,GS08,GMS09,Rez09,bartlett07,Massar95,Gisin99,PeresScudo01,Bagan00, bartlett09,ioannou09,costa09,liang09}. Some of them are directly related to the quantum information science, particularly in the context of quantum communication \cite{bartlett07,Massar95,Gisin99,PeresScudo01,Bagan00, bartlett09,ioannou09} and entanglement detection \cite{bartlett07,costa09,liang09}.

This paper aims to contribute to the aforementioned foundational scenario by formulating the quantum mechanics of spatial degrees of freedom as seen by a {\em reference particle}. As in the Aharonov-Kaufherr (AK) work~\cite{aharonov84}, the idea consists in abandoning the primitive need of well-localized labs of infinite mass to support the theory. In particular, the problem we are concerned with may be further elaborated as follows. Usually, our quantum description of a dynamical system starts by modeling a physical potential and an initial quantum state, both of which are defined relatively to a {\em fixed} system of coordinates in the laboratory. Strictly speaking, therefore, this scheme holds only for {\em inertial} laboratories~\cite{note1}. From the perspective of an external observer, such a lab would be a spatially localized physical system with infinite mass, so as not to get any kick-back under interactions with the particles observed inside. Since there does not exist such an idealized frame in the universe, it is rather legitimate to ask what the physical laws are relative to realistic finite-mass objects. This formulation would automatically lead us to the laws of {\em noninertial quantum frames}, within which fictitious potentials should manifest. Moreover, the very basic notion of reference frame would be suitably accommodated within the conceptual framework of the quantum theory.

Our approach is similar in spirit to Aharonov and Kaufherr's in that we start with the quantum description of a many-particle system from the viewpoint of an {\em absolute} external coordinate system, to be lately abandoned. We then move to a description relative to one of the quantum particles of the system---the quantum frame of reference. However, our approach is still different in three important aspects. First, we employ canonical transformations which provide information about the relative momenta as well. As we show, this allows for the correct physical interpretation of the relational correlations appearing in the relative states. Second, we do consider that the reference particle interacts with other parts of the system, situation which will naturally offer a description from the perspective of accelerated frames of reference. Third, and more important, we discuss several fundamental aspects associated with the relativity of quantum states. Yet, we should stress that our contribution constitutes both a substantial complement and a fundamental extension to studies recently reported by one of us and collaborators~\cite{angelo11} on the physics seen from the perspective of very light frames of reference. 

This paper is organized as follows. In Sec.~\ref{kinematics} we present the main aspects underlying the kinematics of quantum frames of reference in systems of distinguishable particles and examine some fundamental elements which complement and generalize those reported in Ref.~\cite{angelo11}. In particular, we show that no quantum reference frame is disprivileged in relation to absolute frames. Importantly, we identify the conditions which reproduce the inertial description, thus finding out, from the quantum substratum, the defining characteristics of the absolute frame of reference. In Sec.~\ref{dynamics} we present the Hamiltonian formulation of noninertial quantum frames of reference and identify how the fictitious forces formally manifest. Closing remarks are left to Sec.~\ref{conclusion}.

\section{Kinematics}
\label{kinematics}

In this section we elaborate on the kinematics of quantum reference frames. Our strategy consists in starting with the description of the quantum state of a many-particle system from the viewpoint of an absolute frame of reference ({\em external} to the physical system) and then moving to a description relative to one of the particles of the system (an {\em internal reference frame}). 

For simplicity, we initially consider a system with only two particles, say particle 0 and particle 1. Throughout this paper we always use the index 0 to name the particle which is going to be promoted to reference frame. Physically, it is instructive to imagine that particle 0 is a very light closed lab, with no access to the external world, and particle 1 is the system of interest. A central ingredient of the approach is the expression
\be
\ket{x_0}_0\ket{x_1}_1=\ket{\frac{m_0x_0+m_1x_1}{M}}_{\cm}\ket{x_1-x_0}_{r_1},
\label{map}
\ee
which establishes the change of basis connecting two distinct coordinate systems. Here $M=m_0+m_1$, the total mass of the system. The ket $\ket{x_k}_k$, $k=0,1$, denotes the position eigenstate of particle $k$ relative to the external frame of reference, {\em inertial} by hypothesis. The first ket on the r.h.s. is the eigenstate of the center-of-mass position, which is also defined relatively to the external frame. The ket $\ket{x_1-x_0}_{r_1}$, by its turn, encapsulates the physics we want to investigate, as it is a position eigenstate of particle 1 relative to particle 0. Relation \eqref{map} tells us how to express a generic absolute state
\be
\ket{\psi}=\int dx_0 dx_1 \,\psi(x_0,x_1)\, \ket{x_0}_0\ket{x_1}_1
\label{psi2p}
\ee
in terms of the new coordinates. Elementary manipulations yield
\be
\ket{\psi}= \int  dx_{\cm}dx_{r_1} \,\tilde{\psi}(x_{\cm},x_{r_1})\, \ket{x_{\cm}}_{\cm}\ket{x_{r_1}}_{r_1},
\ee
where $\tilde{\psi}(x_{\cm},x_{r_1})=\psi\left(x_{\cm}-\textrm{\scriptsize$\frac{m_1}{M}$}x_{r_1},x_{\cm}+\textrm{\scriptsize$\frac{m_0}{M}$}x_{r_1}\right)$.
Since the center-of-mass position is information defined relatively to the external frame of reference and, therefore, cannot be accessed from inside the closed lab, its corresponding partition must be traced out. Thus, the physics seen from the quantum frame must be revealed solely by the reduced matrix $\hat{\rho}_{r_1}=~\text{Tr}_{\cm}(\ket{\psi}\bra{\psi})$. For convenience, we replace $(x_{\cm},x_{r_1})$ with $(u+m_1\chi/M,\chi)$, where $u$ is a new dummy variable and $\chi$ the relative coordinate. Taking $\delta$ for an arbitrary displacement, the resulting matrix elements of the relative state read
\be
\bra{\chi}\hat{\rho}_{r_1}\ket{\chi+\delta}=\int du~e^{u\partial_{\chi}}\Big[\psi\left(u,\chi\right) \psi^*\hskip-0.1cm\left(u-d_0,\chi+d_1 \right)\Big],
\label{coherences}
\ee
where $d_0=m_1\delta/M$, $d_1=m_0\delta/M$, and $e^{u\partial_{\chi}} f\left(\chi\right)= f\left(\chi+u\right)$. Diagonal elements are given by
\be
\bra{\chi}\hat{\rho}_{r_1}\ket{\chi}=\int du~e^{u\partial_{\chi}} \left|\psi\left(u,\chi\right)\right|^2.
\label{populations}
\ee
In particular, if $\psi(x_0,x_1)=\phi_0(x_0)\phi_1(x_1)$ one has that
\be
\bra{\chi}\hat{\rho}_{r_1}\ket{\chi}=\left[\int du~\left|\phi_0(u)\right|^2 \,e^{u\partial_{\chi}}\right]\left|\phi_1(\chi)\right|^2,
\ee
from which one notes that the relative probability distribution is given by the convolution of the two probability distributions seen by the external observer, meaning that the physics seen by particle 0 is blurred by its own dispersion relative to the external absolute frame. The shift operator $e^{u\partial_{\chi}}$ expresses the relational correction in the position of particle 1 relative to particle 0.

Now, note that the probability distribution $\bra{\chi}\hat{\rho}_{r_1}\ket{\chi}$ seen by particle 1 is always insensitive to the masses. No surprise so far; this is what one would expect from a purely classical relational reasoning. Indeed, it is straightforward to show, under a corresponding change of variables and posterior integration of $x_{\cm}$, that a generic classical distribution $\rho(x_0,x_1)$ produces $\rho_{r_1}(\chi)=\int du\, e^{u\partial_{\chi}}\rho(u,\chi)$, in direct reference to Eq.~\eqref{populations}. On the other hand, Eq.~\eqref{coherences} reveals an important aspect of the physics seen from quantum frames. Actually, it illustrates, for a generic state, a subtle point behind the puzzles investigated in Ref.~\cite{angelo11}. The nonintuitive element comes from the dependence on the masses. Why does the ratio between the masses, which $d_0$ and $d_1$ depend on, influence the relative physics? The answer advanced in Ref.~\cite{angelo11}, and here generalized by Eq.~\eqref{coherences}, points to the fact that the center of mass is typically correlated with the relative partition. Indeed, this is expected to be the case whenever the state is prepared in the external frame; from within the quantum frame, the center of mass cannot be touched and hence can never get correlated with the relative (internal) partition. In \ref{comparison_AK} we show how these aspects manifest in the Aharonov-Kaufherr approach, in which the canonical transformation does not depend explicitly on the masses.

Another important information can be extracted from the result~\eqref{coherences}. Let us assume that $m_1/m_0\to 0$. In this limit we expect the quantum frame to be inertial and, as such, equivalent to the external frame. We then ask whether the internal and external predictions agree. Noting by Eq.~\eqref{coherences} that $d_0\to 0$ and $d_1\to \delta$ we can write
\be
\bra{\chi}\hat{\rho}_{r_1}\ket{\chi+\delta}=\int du~e^{u\partial_{\chi}}\Big[\psi\left(u,\chi\right) \psi^*\hskip-0.1cm\left(u,\chi+\delta \right)\Big].  
\label{rhor1}
\ee
In relation to the external frame, it is straightforward to show, from the state \eqref{psi2p}, that the reduced state of particle 1, $\hat{\rho}_1=\text{Tr}_0\left(\ket{\psi}\bra{\psi}\right)$, has matrix elements
\be
\bra{\chi}\hat{\rho}_{1}\ket{\chi+\delta}=\int du~\psi\left(u,\chi\right) \psi^*\hskip-0.1cm\left(u,\chi+\delta \right).  
\label{rho1}
\ee
Clearly, the relative description \eqref{rhor1} is not equal to the external one yet: the former is an average over all possible displacements promoted by the shift operator $e^{u\partial_{\chi}}$. As shown previously, this introduces additional dispersion to the relative description. Let us assume, in addition, that there is {\em no entanglement}. Considering that $\psi(x_0,x_1)=\phi_0(x_0)\phi_1(x_1)$ makes Eq.~\eqref{rho1} reduce to $\bra{\chi}\hat{\rho}_{1}\ket{\chi+\delta}=\phi_1(\chi)\phi_1^*(\chi+\delta)$, whereas
\be
\bra{\chi}\hat{\rho}_{r_1}\ket{\chi+\delta}=\left[\int du~\left|\phi_0(u)\right|^2~e^{u\partial_{\chi}}\right]~\phi_1(\chi)\phi_1^*(\chi+\delta).
\ee
The descriptions are still different. However, from the above equation we identify a third condition, which finally ensures equivalence, namely, that the state of particle 0 should be highly localized. In fact, by taking $|\phi_0(x_0)|^2=\delta(x_0-\bar{x}_0)$ one obtains that
\be
\bra{\chi}\hat{\rho}_{r_1}\ket{\chi+\delta}=e^{\bar{x}_0\partial_{\chi}}~\phi_1(\chi)\phi_1^*(\chi+\delta),
\ee
which is identical to the external description up to an expected shift in the coordinate system. We have seen, therefore, that the quantum frame becomes equivalent to the external one as long as (i) it is much heavier than the particle under investigation, (ii) it is not entangled with this particle, and (iii) its state is sharply localized in relation to the external frame.

This analysis suggests that these are the aspects tacitly assumed for our everyday laboratories, within which the ordinary quantum-mechanical formulation applies. Remarkably, it reinforces the notion of classicality often associated with these labs and, more importantly, qualifies this classical limit departing from the quantum substratum. Indeed, one may note by $\Delta x_0\Delta p_0=m_0\Delta x_0\Delta \dot{x}_0\geqslant \hbar/2$ that because $m_0\to \infty$ no increase in the speed variance is required as $\Delta x_0\to 0$. Then, low dispersions in position and speed are simultaneously allowed without any violation of the uncertainty principle.

Before closing this section, it is worth showing how our approach connects with the general theory of quantum reference frames~\cite{bartlett07}. Let us consider a unitary operator,
\be
\hat{T}:=\exp\left(-\frac{i \hat{x}_1\hat{p}_0}{\hbar}\frac{m_1}{M}\right)~\exp\left(\frac{i\hat{x}_0\hat{p}_1}{\hbar}\right),
\label{T}
\ee
which yields 
\be
\hat{T}\ket{x_0}_0\ket{x_1}_1=\ket{\frac{m_0x_0+m_1x_1}{M}}_0\ket{x_1-x_0}_1.
\label{Tkets}
\ee
Although the mathematical result is identical to~\eqref{map}, the interpretation is slightly different. Here we have an {\em active} transformation which changes the state of the system while keeping the same coordinate system. The reduced density operator $\hat{\rho}_1=~\text{Tr}_0(\hat{T}\ket{\psi}\bra{\psi}\hat{T}^{\dag})$ can be written
\be
\hat{\rho}_{1}&=&\int du\,dx_1\,dx_1'~\psi\left(u-\frac{m_1}{M}x_1,u+\frac{m_0}{M}x_1\right) \nonumber \\ &\times&\psi^*\left(u-\frac{m_1}{M}x_1',u+\frac{m_0}{M}x_1'\right)~\ket{x_1}\bra{x_1'}.
\label{hatrho1}
\ee
It is straightforward to show that the matrix elements $\bra{\chi}\hat{\rho}_1\ket{\chi+\delta}$ are equal to those given by Eq.~\eqref{coherences}, thus proving that our approach is mathematically equivalent to an active unitary transformation. Now, noting that
\be
\psi\left(u-\frac{m_1}{M}x_1,u+\frac{m_0}{M}x_1\right)=\bra{x_1}\,\hat{T}_u(u)\,\langle u\ket{\psi},
\ee
with
\be
\hat{T}(u)=e^{-\frac{m_1}{M}\hat{x}_1\partial_{u}}e^{\frac{i u\hat{p}_1}{\hbar}},
\ee
we may rewrite Eq.~\eqref{hatrho1} in the form
\be
\hat{\rho}_1=\int du~\hat{T}(u)~\bra{u}\hat{\rho}\ket{u}~\hat{T}^{\dag}(u).
\ee
This formula makes direct reference to the {\em twirling operation}~\cite{bartlett07}. The integration over all possible values of the particle-0 position, which according to Eq.~\eqref{Tkets} plays the role of the center-of-mass position before the active transformation, ensures the invariance of the resulting relative state under translations of the center of mass. This completes the proof that our approach formally enters the general theory of quantum reference frames in the context of an active unitary transformation.

\subsection{Many-particle systems}

A generalization of the above ideas is given as follows. For one-dimensional systems composed of $N+1$ distinguishable particles, a generic state can be written in terms of the eigenstates of external positions as
\be
|\psi\rangle = \int dx_0 ~d^N\vec{x}~\psi(x_0,x_j)~
\ket{x_0}\ket{x_j},
\label{gs}
\ee
where $\ket{x_0}\ket{x_j}\equiv |x_0\rangle |x_1\rangle \ldots |x_N\rangle$, $\psi(x_0,x_j)= \bra{x_0}\bra{x_j}\psi\rangle$, and $d^N\vec{x}=dx_1\ldots dx_N$. Now we move to the description relative to particle 0. We consider a new set of $N+1$ coordinates,
\be\begin{array}{l}
\displaystyle x_{\cm}=\frac{1}{M}\sum_{i=0}^N m_i\,x_i,\\ 
x_{r_j}=x_j - x_0 \qquad (j=1,\ldots,N),
\end{array}
\label{mapN}
\ee
$M=\sum\limits_{i=0}^Nm_i$, whose inverse transformation reads
\be
x_{j}=x_{r_j}+x_{\cm}-\frac{1}{M}\sum_{k=1}^{N}m_k \, x_{r_k},
\label{xi}
\ee
for $j=1,\ldots,N$. Accordingly, we use the generalization of the mapping \eqref{map} to rewrite Eq.~(\ref{gs}) as
\be
|\psi\rangle = \int dx_{\cm} ~d^N\vec{x}_r~\tilde\psi(x_{\cm},x_{r_j})~
\ket{x_{\cm}}\ket{x_{r_j}},
\label{gsn}
\ee
where $\vec{x}_r=(x_{r_1},\ldots,x_{r_N})$ and $\tilde\psi(x_{\cm},x_{r_j})= \psi(x_{0},x_j)$ with $x_0$ and $x_j$ given by Eq.~\eqref{xi}. Tracing out the center-of-mass degree of freedom we finally get
\be
\bra{\chi_j}\hat\rho_{r}\ket{\chi_j+\delta_j} =
\int du~e^{\vec{u}\cdot\nabla_{\chi}}\Big[\psi(u,\chi_j)
\psi^*(u-\Delta,\chi_j-\Delta+\delta_j)
\Big],
\label{rhor}
\ee
and
\be
\bra{\chi_j}\hat\rho_{r}\ket{\chi_j} =
\int du ~e^{\vec{u}\cdot\nabla_{\chi}}\left|\psi(u,\chi_j)\right|^2,
\ee
where $\vec{u}\cdot\nabla_{\chi}=u\sum_j\partial_{\chi_j}$ and $\Delta=\sum_{j=1}^Nm_j\delta_j/M$. It is just an exercise to prove that, despite minor differences in the formulas, all the conclusions of the two-particle case directly apply here. The corresponding many-particle active transformation is given by
\be
\hat{T}=\exp\left(-\frac{i \hat{p}_0}{\hbar}\sum\limits_{j=1}^N\frac{m_j\hat{x}_j}{M}\right)
\exp\left(\frac{i\hat{x}_0}{\hbar}\sum\limits_{j=1}^N\hat{p}_j\right).
\ee
%

\subsection{Relative nonlocality} \label{rnonlocality}

Another important aspect of quantum reference frames appears specifically in situations involving more than two particles. It refers to the structure of the Hilbert space. We consider two distinct canonical transformations involving relative coordinates, both illustrated bellow for $N=~3$:
\be\label{xrpi3}
&&\begin{array}{lll}
\hat{x}_{\cm} = \frac{1}{M}\sum\limits_{i=0}^2m_i\hat{x}_i,&\quad & \hat{\pi}_{\cm}=\sum\limits_{i=0}^2\hat{p}_i= \hat{p}_{\cm},\\ \\
\hat{x}_{r_1} = \hat{x}_1-\hat{x}_0,&\quad &\hat{\pi}_1=m_1\left(\frac{\hat{p}_1}{m_1}-\frac{\hat{p}_{\cm}}{M}\right),  \\ \\
\hat{x}_{r_2} = \hat{x}_2-\hat{x}_0,&\quad & \hat{\pi}_{2}=m_2\left(\frac{\hat{p}_2}{m_2}-\frac{\hat{p}_{\cm}}{M}\right),
\end{array} 
\ee
and
\be\label{qpr3}
&&\begin{array}{lll}
\hat{q}_{\cm}=\frac{1}{M}\sum\limits_{i=0}^2m_i\hat{x}_i=\hat{x}_{\cm},&\;\;&\hat{p}_{\cm} = \sum\limits_{i=0}^N\hat{p}_i, \\ \\
\hat{q}_1=\gamma\left(\hat{x}_1-\frac{m_0\hat{x}_0+m_2\hat{x}_2}{m_0+m_2}\right),& & \hat{p}_{r_1}= \mu_{01}\left(\frac{\hat{p}_1}{m_1}-\frac{\hat{p}_0}{m_0}\right), \\ \\
\hat{q}_2=\gamma\left(\hat{x}_2-\frac{m_0\hat{x}_0+m_1\hat{x}_1}{m_0+m_1}\right),& & \hat{p}_{r_2} = \mu_{02}\left(\frac{\hat{p}_2}{m_2}-\frac{\hat{p}_0}{m_0}\right),
\end{array} \qquad
\ee
where $\gamma=\frac{m_0 m_1 m_2}{M\mu_{01}\mu_{02}}$ and $\mu_{0j}=m_0m_j/(m_0+m_j)$. It is immediately seen that the relative observables accessible from within the quantum reference frame, namely $\hat{x}_{r_j}$ and $\hat{p}_{r_j}$, are {\em not} canonically conjugated operators, so that they do not span a joint Hilbert space in the usual sense. Actually, the Hilbert space spanned by them are intrinsically nonlocal in the sense that
\be
e^{-i\,\delta\, \hat{p}_{r_1}/\hbar}\ket{x_{\cm}}\ket{x_{r_1}}\ket{x_{r_2}}=\ket{x_{\cm}}\ket{x_{r_1}+\delta}\ket{x_{r_2}+\frac{\mu_{01}}{m_0}\delta}.
\ee
(We omit the subindexes ``$\cm$'' and ``$r_j$'' in the kets when there is no risk of confusion.) As shown in details in Ref.~\cite{angelo11}, via analysis of a paradox, this result is the expression of the kickback received by particle 0 in shifting particle 1; the resulting back-action in the frame implies a relative shift of particle 2 as well. The analogy with classical physics accurately applies: when we push the planet downwards we get a kickback upwards and then we see all the remote stars moving together with the planet. In the limit $m_0\to~ \infty$ this relative nonlocality disappears, as expected. Indeed, this can be verified by a direct inspection of the above formulas.

\subsection{Uncertainty relations and information}\label{uncertainty}

For arbitrary observables $\hat{A}$ and $\hat{B}$ the strongest form of the uncertainty relation reads $\Delta A\,\Delta B\geqslant \frac{1}{2} |[\hat{A},\hat{B}]|$, where $\Delta A$ $(\Delta B)$ stands for the variance of $\hat{A}$ ($\hat{B}$). Naturally, one may ask how the uncertainty relations look from the perspective of a quantum reference frame. Let us consider the extension of the transformation \eqref{xrpi3} to arbitrary $N$,
\be\label{xrpiN}
\begin{array}{lll}
\hat{x}_{\cm}=\frac{1}{M}\sum\limits_{i=0}^Nm_i\hat{x}_i,&\quad & \hat{\pi}_{\cm}=\sum\limits_{i=0}^N\hat{p}_i=\hat{p}_{\cm},\\ \\
\hat{x}_{r_j}=\hat{x}_j-\hat{x}_0,&\quad &\hat{\pi}_j=m_j\left(\frac{\hat{p}_j}{m_j}-\frac{\hat{p}_{\cm}}{M} \right),
\end{array}
\ee
where particle 0 is again the quantum reference frame. The momentum of the $j$-th particle relative to particle 0 is given by $\hat{p}_{r_j}=~\mu_{0j}\left(\frac{\hat{p}_j}{m_j}-\frac{\hat{p}_0}{m_0}\right)$. By computing the commutation relations among all the relative operators one sees that the only unusual uncertainty relation is given by
\be
\Delta x_{r_j}\,\Delta p_{r_k}\geqslant\frac{\hbar}{2}\left(\frac{m_k}{m_0+m_k}\right) \qquad (j\neq k).
\ee
This is another expression of the relative nonlocal effect associated with the lightness of the quantum frame. In fact, if $m_0\to\infty$ one gets the usual relation, i.e., $[\hat{x}_{r_j},\hat{p}_{r_k}]=0$ and $\Delta x_{r_j}\,\Delta p_{r_k}\geqslant 0$~\cite{note2}. Further interesting information may be obtained from the l.h.s. of the above inequality. Using Eq.~\eqref{xrpiN} and the definition of variance one shows that
\be
\begin{array}{l}
\left(\Delta x_{r_j}\right)^2=\left(\Delta x_0\right)^2+\left(\Delta x_j\right)^2-2\,C(x_j,x_0),\\ \\
\left(\Delta p_{r_j}\right)^2=\frac{\mu_{0j}^2}{m_0^2}\left(\Delta p_0\right)^2+\frac{\mu_{0j}^2}{m_j^2}\left(\Delta p_j\right)^2-2\,\mu_{0j}^2\frac{C(p_j,p_0)}{m_jm_0},
\end{array}
\ee
where $C(A,B):=\langle\frac{\hat{A}\hat{B}+\hat{B}\hat{A}}{2}\rangle-\langle\hat{A}\rangle\langle\hat{B}\rangle$. As we have seen in a previous analysis, the relative variances of the $j$-th particle is increased by the external dispersions $\Delta x_0$ and $\Delta p_0$ of the quantum frame. Now, however, we note that some correlations, as denounced by $C(x_j,x_0)$ and $C(p_j,p_0)$, may diminish the relative dispersions thus increasing the relative information about $x_{r_j}$ and $p_{r_j}$. This is in agreement with the common-sense intuition according to which a correlation between two parts implies that information about the state of one is available to the other. Two aspects underlying these ideas can be illustrated by the following simple example. 

Consider two particles of equal masses in a superposition state $\ket{x_0}\ket{x_1}+\ket{x_0+\delta}\ket{x_1+\delta}$, with a proper normalization, composed of Gaussian states $\ket{x}$ centered at $x$ and with variance $\Delta$. Given that $\langle x_{0,1}|x_{0,1}+\delta\rangle=e^{-\alpha}$, where $\alpha\equiv\delta^2/8\Delta^2$, one shows that the purity of the subsystem is given by $\mathcal{P}(\hat{\rho}_1)=\text{Tr}_1\hat{\rho}_1^2=1-\frac{1}{2}\tanh^2(\alpha)$. In particular, for well-separated branches in the superposition state, i.e., for $\alpha\gg 1$, the purity approximates $1/2$, indicating that the reduced state of particle 1 relative to the external frame is almost maximally mixed. In this sense, practically no information about partition 1 is available to the external reference frame. On the other hand, in the new basis, the state is written $(\ket{x_{\cm}}+\ket{x_{\cm}+\delta})\ket{x_1-x_0}$, so that particle 1 is in a pure state from the perspective of the quantum frame for any $\delta$; all the information is available in there. The second aspect can be appreciated by comparing the variances $\Delta x_{r_1}=\sqrt{2}\Delta$ and
\be
\Delta x_1=\Delta x_{r_1}\sqrt{\frac{1}{2}+\frac{\alpha}{2}\Big[1+\tanh(\alpha)\Big]}.
\ee
For $\alpha\gg 1$, one sees that $\frac{\Delta x_1}{\Delta x_{r_1}}\approx \sqrt{\alpha}\gg 1$, i.e., much more information about the position of particle 1 is available to the quantum frame, the one correlated with the particle. In fact, it is interesting to see how the ratio between the dispersions behaves as a parametric function of entanglement. For the state under consideration, the entanglement is given, via linear entropy, by $E=1-\mathcal{P}(\hat{\rho}_1)=\frac{1}{2}\tanh^2(\alpha)$. It follows that
\be
\frac{\Delta x_1}{\Delta x_{r_1}}=\sqrt{\frac{1}{2}+\frac{(1+\sqrt{2E})\,\text{arctanh}(\sqrt{2E})}{2}},
\ee
a monotonically increasing function of $E$. Therefore, the more entangled the state from the viewpoint of the external frame the less the dispersion of the relative position in comparison with the external dispersion and, hence, the more the knowledge of particle 0 on the position of particle 1 as compared to the knowledge available to the external frame. In particular, when entanglement is maximum one has that $\frac{\Delta x_1}{\Delta x_{r_1}}\to\infty$. It is possible to recognize from this quantitative analysis some elements which support the motivating issues underlying Everett's work~\cite{everett57}.

So far we have pointed out several conceptual aspects of quantum reference frames but have not provided any discussion about physical realizations of such ideas. Indeed, one may wonder to which extent this scenario would actually manifest in practice. In the next sections, we pursue this aim through the analysis of simple instances involving a few particles.

\subsection{Measuring phases}
\label{phases}

Consider the following state of two particles:
\be
\ket{\psi}=\frac{1}{\sqrt{2}}\Big(\ket{x_0}\ket{x_1}+e^{i\phi}\ket{x_0+\delta_0}\ket{x_1+\delta_1} \Big).
\label{psi}
\ee
Again $\ket{x_i}$ denotes a Gaussian state with dispersion $\Delta_i\ll~\delta_{i}$ and center at position $x_i$ relative to the external frame. Using the mapping \eqref{map} we obtain
\be
\ket{\psi}=\frac{1}{\sqrt{2}}\Big(\ket{x_{\cm},x_{r_1}}+e^{i\phi}\ket{x_{\cm}+\delta_{\cm},x_{r_1}+\delta_{r_1}} \Big),\quad
\label{int_ent}
\ee
where 
\be
\delta_{\cm}=\frac{m_0\delta_0+m_1\delta_1}{m_0+m_1} \quad\text{and}\quad  \delta_{r_1}=\delta_1-\delta_0.
\ee
A remark is in order. The state $\ket{x_{\cm},x_{r_1}}$ is, so to speak, an {\em internally entangled} state, i.e., it cannot be written as $\ket{x_{\cm}}\ket{x_{r_1}}$ in general. Unlike the situation involving position eigenstates, such as in Eq.~\eqref{map}, these {\em internal correlations} naturally manifest in general due to the change of basis. A direct calculation of the purity of the relative state, $\hat{\rho}_r=\textrm{Tr}_{\cm}\ket{x_{\cm},x_{r_1}}\bra{x_{\cm},x_{r_1}}$, shows that
\be
\mathcal{P}(\hat{\rho}_r)=\textrm{Tr}_r\hat{\rho}_r^2=\frac{(m_0+m_1)\Delta_0\Delta_1}{\left[(m_0^2\Delta_0^2+m_1^2\Delta_1^2)(\Delta_0^2+\Delta_1^2)\right]^{1/2}},\qquad
\ee
formula which can be directly generalized for many-particle Gaussian states of form $\ket{x_0}\ket{x_1}\cdots\ket{x_N}$.
One sees that $\ket{x_{\cm},x_{r_1}}$ separates into $\ket{x_{\cm}}\ket{x_{r_1}}$ only via a perverse choice of parameters, namely, $m_0\Delta_0^2=m_1\Delta_1^2$. Off this regime, such an internal entanglement will persist. It is important to ponderate, therefore, what the implications of such internal entanglement are for an observer within the quantum frame. One may ask, for instance, whether the phase in Eq.~\eqref{int_ent} could be accessed in practice from within the quantum frame when $\delta_{\cm}=0$, i.e., when entanglement is present only internally.

To answer this question, we employ the approach advanced in Ref.~\cite{aharonov05}, which consists in considering the mean value of the shift operator $e^{-i\hat{S}/\hbar}$, where $\hat{S}=~\delta_0\hat{p}_0+\delta_1\hat{p}_1$. Using the state \eqref{psi} we obtain that
\be
\bra{\psi}e^{i\hat{S}/\hbar}\ket{\psi}=e^{i\phi}.
\ee
Experimentally, the evaluation of the phase $\phi$ would require the external observer to shift both particles by the respective distances $\delta_0$ and $\delta_1$. Presumably, the procedure involves to make two copies of the state $\ket{\psi}$ (e.g., using a beam splitter), shift one of them so as to get $e^{i\hat{S}/\hbar}\ket{\psi}$, and then look at the overlap with the unshifted one, $\ket{\psi}$. 

In terms of the new set of operators,
\be
\hat{S}=\delta_0\hat{p}_0+\delta_1\hat{p}_1=\delta_{\cm}\hat{p}_{\cm}+\delta_{r_1}\hat{p}_{r_1}.
\label{S}
\ee
Now we see, from a different perspective, that in general it is necessary to shift not only particle 1 relatively to particle 0 but also the center of mass relatively to the external frame. But the position of the center of mass cannot be changed from within the quantum frame, which, in addition, has been assumed from the outset to have no access to the outside world. Then, whenever $\delta_{\cm}\neq 0$ the phase cannot be accessed by particle 0. Likewise, such a state could not even be generated within the quantum frame. Therefore, without an external reference frame this kind of state would be {\em forbidden} to exist---an illustration, in terms of spatial degrees of freedom, of the link between a superselection rule and the lack of a quantum reference frame~\cite{aharonov155,aharonov158,bartlett07}. On the other hand, if $\delta_{\cm}=0$, then $\hat{S}=\delta_{r_1}\hat{p}_{r_1}$ and $e^{i\hat{S}/\hbar}\ket{x_{\cm},x_{r_1}}=\ket{x_{\cm},x_{r_1}-\delta_{r_1}}$, so that the phase can be accessed via internal interactions only, despite the existence of internal entanglement in state \eqref{int_ent}. 

We have advanced towards a more realistic scenario by identifying an operator whose mean value is able to diagnose whether relative superpositions can be physically detected from inside the quantum reference frame. Next, we move one step forward. Further realistic elements are discussed which illustrate the ideas presented thus far and unveil another important aspect underlying the theory of quantum frames of reference.

\subsection{No disprivileged quantum frame of reference}

Consider a double-slit experiment to be realized inside a light rocket. When the particle passes through the upper (lower) slit, it transfers linear momentum to the rocket which then moves upwards (downwards). The information about the motion of the rocket is available only to external observers; it cannot be accessed from within the closed rocket. The situation, therefore, is such that the position of the rocket provides {\em which-way information} which could be used by external observers to find out the path chosen by the particle in the experiment. Since which-way information is not available inside the rocket one may wonder whether the external reference frame is somehow {\em privileged} in relation to the internal one. This thought experiment seems to defy Bohr's complementarity principle in the following sense. In possession of which-way information, an external observer would access the corpusclelike behavior of the particle whereas for an internal observer the wavelike behavior should manifest. After all, what would be the outcome of such an experiment?

\begin{figure}[ht]
\centerline{\includegraphics[scale=0.5]{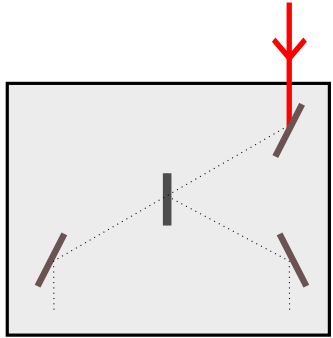}}
\caption{A particle impinges on a board which is allowed to freely move. The dotted lines denote the two possible paths in a situation in which the board is much heavier than the particle.}
\label{setup}
\end{figure}

We explore this issue by analyzing a simplified version of this experiment (see Fig.~\ref{setup}). Consider an isolated system of two quantum objects. One of them, a particle of mass $m_p$, moves towards the other, a very light board of mass $m_b$. A beam splitter and three mirrors are rigidly attached to the board, which is allowed to freely move in space.

Before the particle reaches the first mirror we assume that the state of the system is given by a product of sharp Gaussian states, $\ket{0}_b\ket{L}_p$, centered at the positions $0$ and $L$ relative to some external coordinate system. After passing the beam splitter and the two later mirrors the particle gets entangled with the board. This occurs because the board moves in virtue of the momentum conservation. The state of the system then reads
\be
\ket{\psi}=\frac{1}{\sqrt{2}}\left(\ket{d}\ket{-L+d}+e^{i\pi/2}\ket{0}\ket{L}\right)_{b,p},\quad
\label{psibp}
\ee
where $d= 2L m_p/(m_p+m_b)$.
The term $e^{i\pi/2}$ accounts for the phase difference introduced by the beam splitter between reflexion and transmission. Notice that if $m_b\to\infty$, then $d\to 0$ and the situation often observed in our everyday laboratories, with no entanglement between the particle and the board, is recovered. However, for comparable masses, entanglement persists and the reduced state for the particle no longer retains information about the phase. As mentioned above, which-way information is available to the external frame so that the particle cannot manifest wavelike behavior.

The physics seen by the board can be assessed by rewriting state \eqref{psibp} in terms of the new coordinates:
\be
\ket{\psi}&=&\frac{1}{\sqrt{2}}\left(\ket{\frac{d}{2},-L}+e^{i\pi/2}\ket{\frac{d}{2},L}\right)_{\cm,r_p}.\label{psibp_cmr}
\ee
Here no entanglement exists except the internal one, which has been shown not to prevent the observation of the phase. Therefore, one may conclude that the board, without any which-way information about itself, sees the particle in a superposition state, which is representative of the wavelike behavior. Thus, the contrast between the external and internal predictions has been made explicit.

\begin{figure}[htb]
\centerline{\includegraphics[scale=0.4]{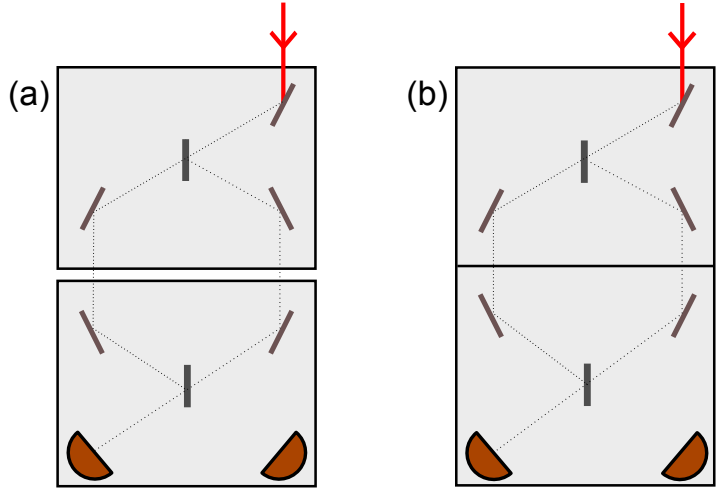}}
\caption{A measuring device, composed of two mirrors, a beam splitter, and two detectors, is rigidly attached either (a) to the external frame or (b) to the board. Dotted lines represent the possible paths for the case in which the whole setup is much heavier than the impinging particle.}
\label{interferometer}
\end{figure}

The solution to the conflict emerges when we explicitly consider the measurement process. This implies taking into account the spatial state of the measuring device, clearly specifying to which reference frame it is attached. Furthermore, as we show next, the correct prediction is made by asking how the state looks relatively to the measuring device. The aim of this strategy is to take the discussion to the level of the measurement, thus avoiding nonphysical preconceptions. We introduce the measuring device (MD) in the ways depicted in Fig.~\ref{interferometer}. 

In situation (a) the MD is outside the quantum frame. In order for the MD to assess the predictions of an external observer it cannot move relatively to this observer. As we have learned at the beginning of Sec.~\ref{kinematics}, this requires the MD to possess an apparent mass $m_{\MD}\to \infty$. Besides, the MD must be in a sharply localized state relative to the external frame, so as to ensure that its position is effectively known. After the interaction of the particle with the board, the state of the system is $\ket{\psi_{(a)}}=\ket{0}_{\MD}\ket{\psi}$, where $\ket{0}_{\MD}$ denotes that the MD is well localized at the position 0 (relative to the external frame) and $\ket{\psi}$ is given by Eq.~\eqref{psibp}. Now, in terms of coordinates relative to the MD one has that
\be
\ket{\psi_{(a)}}=\frac{1}{\sqrt{2}}\left(\ket{0,d,-L+d}+e^{i\pi/2}\ket{0,0,L}\right)_{\cm,r_b,r_p},
\label{psi(a)}
\ee
in agreement with the predictions of the external observer as derived from Eq.~\eqref{psibp}. Note that the center-of-mass position is precisely in the MD location. Definitely, the MD will receive the particle in a mixture, as it is entangled with the board. Then, there will be no destructive interference in one of the ports and both detectors will be likely to click. Interestingly, from the perspective of the quantum frame (the board), now the situation is such that the very MD provides which-way information about the particle. 
Indeed, it is not difficult to check that the particle gets entangled with the MD from the perspective of the board:
\be
\ket{\psi_{(a)}}=\frac{1}{\sqrt{2}}\left(\ket{0,-d,-L}+e^{i\pi/2}\ket{0,0,L}\right)_{\cm,r'_{MD},r'_p},
\label{psi(a)fromboard}
\ee
where $r'_{MD}$ and $r'_p$ are, respectively, the coordinates of the MD and the particle seen by the board. Thus, only the corpusclelike character of the particle will manifest to any observer; the conflict dissipates.

On the other hand, in situation (b) the MD is attached to the board and hence can freely move as well. Now the system board+MD turns out to be an interferometer. In this case, after the particle impinges on the interferometer the whole system appears in an entangled state relative to the external frame. In fact, since the MD is rigidly attached to the board the state of the system reads
\be
\ket{\psi_{(b)}}=\frac{1}{\sqrt{2}}\left(\ket{d'}\ket{d'}\ket{-L+d'}+e^{i\pi/2}\ket{0}\ket{0}\ket{L}\right)_{\MD,b,p},
\label{psi(b)}
\ee
where $d'=2L m_p/(m_p+m_b+m_{\MD})$. Moving to the perspective of the MD one finds that
\be
\ket{\psi_{(b)}}=\frac{1}{\sqrt{2}}\left(\ket{\frac{d'}{2},0,-L}+e^{i\pi/2}\ket{\frac{d'}{2},0,L}\right)_{\cm,r_b,r_p}.
\label{psi(b)rel}
\ee
This is in agreement with the prediction of the internal observer as derived from Eq.~\eqref{psibp_cmr}. Actually, the above state precisely gives the physics seen by the board as well, since it is rigidly correlated with the MD. There will be destructive interference in one of the ports; only one of the detectors will click. In this case, however, it seems that the external prediction has not been reconciliated with the internal one yet, since according to Eq.~\eqref{psi(b)} the particle is entangled with the board; which-way information is still available. The conflict resists here because we have not considered the entire dynamics towards the detectors. Actually, Eq.~\eqref{psi(b)} refers to the state after the particle passes by the first part of the interferometer (the board). The complete analysis is crucial in this case because during the measurement process the detectors can move under interactions with the particle. Indeed, by carrying on the calculations, under the assumption of momentum conservation and low dispersion of the Gaussian states, it is straightforward to show that after the particle passes by the last beam splitter the state of the system dynamically evolves to $\ket{d'}_{\MD}\ket{d'}_b\ket{-L+d'}_p$. Therefore, the external which-way information is erased right before the detection. Again, the contradictions are all dissipated.

We are now in position to analyze the puzzling questions concerning the double-slit experiment inside the light closed quantum rocket. When the measuring device is located inside the rocket, the {\em context} has been drawn so as to reveal the internal physics. In analogy to what we have seen above, we may conclude that in such an experiment no which-information is available either internally or externally, so that an interference pattern will be verified inside. On the other hand, when the measuring device is attached to the external frame, the motion of the rocket provides which-way information so that the particle cannot interfere. This is consistent with the viewpoint of the internal observer, since which-way information emerges from the relative motion of the measuring device. Therefore, as far as the theoretical description properly accounts for all physical objects participating in a given experiment---including the measuring device---one may conclude that the quantum reference frame is not {\em disprivileged} in relation to the external frame.

Finally, one should note from the above discussion that the notion of {\em quantum observer} has been properly linked with the measuring device, a finite-mass object prepared in a determined spatial state. In fact, we have seen that the predictions for a given observer, as derived originally from states \eqref{psibp} and \eqref{psibp_cmr}, have materialized only when the measuring device was attached to the frame of the respective observer. Besides being a very healthy strategy from a physical point of view, this approach helps us to formalize some of the concepts explored in this paper and situate them in an experimental context.

\subsection{Paradox of the third particle -- version 2}

The title of this section makes reference to a paradox investigated in Ref.~\cite{angelo11}, in which an apparent innocuous particle is found to influence the relative physics. Here the situation is similar, but the crux of the matter is rather different.

Let us remove the board of the experiment (a) in Fig.~\ref{interferometer}. Now let us assume that a particle of mass $m_p$ approaches the MD in a superposition state. The MD is light and free to move. In addition, we consider that a third particle, of mass $m_3$, is in a sharp state centered at a position $x$ relative to the external frame. No assumption is made on the value of $x$; the third particle may be as far away as one wishes. The state of the system is written
\be
\ket{\psi}=\ket{0}_{\MD}\left(\frac{\ket{-L}_p+e^{i\phi}\ket{L}_p}{\sqrt{2}}\right)\ket{x}_3.
\label{psi3}
\ee
In terms of coordinates relative to the MD one has that
\be
\ket{\psi}=\left(\frac{\ket{\frac{-m_pL+m_3 x}{M},-L,x}+e^{i\phi}\ket{\frac{m_pL+m_3x}{M},L,x}}{\sqrt{2}}\right)_{\hspace{-0.15cm}\cm,r_p,r_3}\hspace{-1.cm},
\label{psi3rel}
\ee
where $M=m_{\MD}+m_p+m_3$. We see that the center of mass is entangled with the relative position of the particle so that the MD cannot access the phase, i.e., both detectors of the MD are likely to click. 

Now, consider that the third particle is much heavier than both the MD and the impinging particle. The state of the system reduces to
\be
\ket{\psi}\simeq\frac{1}{\sqrt{2}}\left(\ket{x,-L,x}+e^{i\pi/2}\ket{x,L,x}\right)_{\cm,r_p,r_3}.
\label{psi_largem3}
\ee
In this regime the entanglement with the center of mass disappears and the phase becomes accessible to the MD. Thus the puzzle emerges: how is it possible that the mass of an arbitrarily distant particle influences the observation of the phase?

Again, it is necessary to discuss how the phase can actually be measured. Based on the formalism of Sec.~\ref{phases} and on the canonical transformations given by Eq.~\eqref{qpr3} one may check that 
\be
\hat{S}=\sum\limits_{i=0}^2\delta x_i\,\hat{p}_i=\delta q_{\cm}\,\hat{p}_{\cm}+\sum\limits_{j=1}^2\delta q_j\,\hat{p}_{r_j}.
\ee
Now, using the state \eqref{psi3} we see that $\bra{\psi}e^{i\hat{S}/\hbar}\ket{\psi}=e^{i\phi}$ if $\hat{S}=2L\,\hat{p}_p$.  According to the above equation, we can write
\be
\hat{S}=2L\left(\frac{m_p}{M}\hat{p}_{cm}+\gamma\hat{p}_{r_p}-\frac{\mu_{\MD\,p}}{m_{\MD}}\gamma\hat{p}_{r_3}\right),
\ee
where $\gamma=m_{\MD}m_pm_3/(M\mu_{\MD p\,} \mu_{\MD 3})$. This result indicates that the phase cannot be accessed by means of internal interactions only; the center of mass has to be shifted. However, if the third particle is sufficiently heavy then
\be
\hat{S}\simeq 2L\left(1+\frac{m_p}{m_{\MD}}\right)\hat{p}_{r_p}-2L\frac{m_p}{m_{\MD}}\hat{p}_{r_3}.\quad
\label{S_prel}
\ee
In this regime only relative shifts are required, so that the phase can be measured. But how does it come in practice?  

First of all, it is noticeable in the above relation that the task can be accomplished only if the third particle is touched. That is, the third particle has to effectively participate in the experiment; it has to be shifted relatively to the quantum frame. However, according to the state \eqref{psi_largem3} only the impinging particle needs to be shifted relatively to the interferometer. Is there any inconsistence here? No. As has been shown in Ref.~\cite{angelo11}, and reinforced here by Eqs.~\eqref{xrpi3} and \eqref{qpr3}, this is just a misleading interpretation emerging from a preconception according to which relative positions and relative momenta are canonically conjugated to each other. This is true only if the frame is much heavier than the system under investigation or in the two-particle case. From Eqs.~\eqref{xrpi3} and \eqref{qpr3} we see that the relative momenta $\hat{p}_{r_p}$ and $\hat{p}_{r_3}$, which span $\hat{S}$ in \eqref{S_prel}, are conjugated to the position operators $\hat{q}_{p}$ and $\hat{q}_{3}$, not to the relative positions $x_{r_p}$ and $x_{r_3}$, to which the state \eqref{psi_largem3} refers. In fact, in terms of the momenta \eqref{xrpi3} we see that
\be
\hat{S}=2L\,\hat{p}_p=2L\left(\frac{m_p}{M}\hat{\pi}_{cm}+\hat{\pi}_p\right)  \simeq 2L\,\hat{\pi}_p,
\label{S_prelcm}
\ee
which now agrees with \eqref{psi_largem3}. The operator $\hat{\pi}_p$ is the momentum of the impinging particle relative to the center of mass. But now the center of mass is precisely at the third particle, which is the heaviest object in the system and does not move relatively to the external frame. In this capacity, the third particle plays the same role of the external frame as is apparent from \eqref{S_prelcm}; the momentum of the impinging particle relative to the center of mass, $\hat{\pi}_p$, is equal to its momentum relative to the external reference frame, $\hat{p}_p$.

Now we see how the phase can actually be observed by the light MD. All one needs to do is to attach the mirrors and the beam splitter to the third particle. The MD no longer needs to shift the particle and just collects it in a superposition state. Once we realize that the third particle is not innocuous at all, the paradox disappears. While the third particle played no physical role in the solution of the first version of the paradox in Ref.~\cite{angelo11} here it emerges as the missing reference frame. To better appreciated this point we should note by Eq.~\eqref{psi3rel} that the center of mass cannot be prepared in an entangled state by means of internal interactions only. The phase $\phi$ has been prepared, therefore, relatively to some external reference frame. This means that in order to access the phase one would have to touch such an external physical structure. However, when particle 3 becomes heavy enough it assumes the role of the external reference frame, so that the phase can be retrieved through internal interactions only.

\section{Dynamics}\label{dynamics}

At last we examine how the dynamics is seen from a quantum reference frame, which, as mentioned before, is naturally {\em noninertial} by conception. Let us consider a closed universe composed of $N+1$ particles interacting via conservative forces. From the point of view of an {\em absolute} reference frame---to be posteriorly abandoned---the total energy of the system can be described by the Hamiltonian 
\begin{eqnarray}\label{H}
\begin{array}{c}
\displaystyle\hat{H}=\sum\limits_{i=0}^N\frac{\hat{p}_i^2}{2m_i}+\hat{V}(\hat{x}_0,\hat{\mathbf{x}}),\\
\displaystyle\hat{V}(\hat{x}_0,\hat{\mathbf{x}})=\sum\limits_{j=1}^NV(\hat{x}_j-\hat{x}_0)+\sum\limits_{j=1}^N\sum\limits_{k>j}^NV(\hat{x}_k-\hat{x}_j),
\end{array}
\end{eqnarray}
where $\hat{\mathbf{x}}=(\hat{x}_1,\dots,\hat{x}_N)$.
The usual formulation of quantum mechanics is assumed to hold relatively to the external absolute frame, i.e., the dynamical evolution of the system is governed by $i\hbar\,\partial_t\ket{\psi}=\hat{H}\ket{\psi}$, with $\hat{H}$ given above. Now we move to the new canonical operators, those given by Eq.~\eqref{xrpiN}. The observable $\hat{x}_{r_j}$ refers to the position of the $j$-th particle relative to particle 0 (the quantum frame) whereas $\hat{\pi}_j$ refers to the momentum of the particle relative to the center of mass. The above Hamiltonian is then rewritten $\hat{H}=\frac{\hat{p}_{\cm}^2}{2M}+\hat{H}_{rel}$, where 
\be\label{Hrel}
\hat{H}_{rel}&=&\sum\limits_{j=1}^N \frac{\hat{\pi}_j^2}{2m_j}+\hat{V}(0,\hat{\mathbf{x}}_r)+\frac{\hat{\Pi}^2}{2m_0},
\ee
with $\hat{\mathbf{x}}_r=(\hat{x}_{r_1},\dots,\hat{x}_{r_N})$, and
\be\label{Lambda}
\hat{\Pi}\equiv \sum\limits_{j=1}^N\hat{\pi}_j=\frac{m_0}{M}\hat{p}_{\cm}-\hat{p}_0.
\ee
Clearly, the center of mass does not interact with the relative degrees of freedom and hence its momentum is constant. Moreover, since there is nothing else outside the system, the center of mass can never get entangled with the relative partitions. It follows that all quantum states prepared via internal interactions will be perceived by particle 0 as a {\em pure} state. On the other hand, from the external perspective, particle 0 will generally be entangled with the other particles. As we have seen previously (Sec.~\ref{uncertainty}), these correlations derive from the fact that in the preparation process the quantum reference frame acquires information about the state of the system. Also, by Eq.~\eqref{Lambda} we see that $\hat{\Pi}/m_0$ can be interpreted as the velocity of particle 0 relative to the center of mass, so that the last term in Eq.~\eqref{Hrel} is the kinetic energy of particle 0 relative to the center of mass. 

The formula we have found for $\hat{H}_{rel}$ differs from AK's~\cite{aharonov84} in that here we allow the particles to interact with the quantum frame. This makes the operator $\hat{\Pi}$---a {\em vector potential} in AK's terminology---to be the expression of noninertial forces acting inside the quantum frame. To verify this we take the time derivative of the Heisenberg velocity,  
\be\label{velocity}
\dot{\hat{x}}_{r_j}=\frac{\hat{\pi}_j}{m_j}+\frac{\hat{\Pi}}{m_0}=\frac{\hat{p}_{r_j}}{\mu_{0j}} \qquad (j\geqslant 1),
\ee
and note, by Eq.~\eqref{Lambda}, that $d\hat{\Pi}/dt=-\dot{\hat{p}}_0=-m_0\ddot{\hat{x}}_0$. We then obtain that
\be\label{acceleration}
\ddot{\hat{x}}_{r_j}=\frac{\left(-\partial_{\hat{x}_{r_j}}\hat{V}\right)}{m_j}-\ddot{\hat{x}}_0=\frac{\dot{\hat{p}}_{r_j}}{\mu_{0j}},
\ee
where $\hat{V}=\hat{V}(0,\hat{\mathbf{x}}_r)$. Clearly, this is the equation of motion of a particle within a {\em noninertial quantum reference frame} that moves with acceleration $\ddot{\hat{x}}_0$ relative to the absolute space. Inside this frame, the quantum dynamics of particle $j$ can be correctly explained only by the introduction of the {\em fictitious force} $\frac{m_j}{m_0}d\hat{\Pi}/dt$.

The operator $\hat{\Pi}$ is, therefore, the formal manifestation of the kick back that the quantum reference frame suffers under interactions with the system. In fact, if particle 0 does not interact with the other particles then $V(\hat{x}_{r_j})=0$ and $i\hbar\,d\hat{\Pi}/dt=[\hat{\Pi},\hat{H}]=0$, which leads \eqref{acceleration} to reproduce the equation of motion expected for an inertial frame. In addition, if $m_0$ is arbitrarily heavy then no recoil is expected for the frame. Accordingly, the kinetic energy $\hat{\Pi}^2/2m_0$ becomes negligible and $\hat{\Pi}$ disappears from the formalism. The typical equations of motion of inertial reference frames are thus readily recovered.

At this point it is insightful to observe an important difference to approaches involving {\em classical frames of reference}. Let us consider a change in the origin of the coordinate system. In classical mechanics, this can be implemented by the transformation $x_j'=x_j-x_0$, which defines the position of the $j$-th particle of the system relatively to an arbitrary, eventually immaterial, {\em classical origin} $x_0$. Usual procedures in classical mechanics allows us to construct a canonical formulation starting from absolute kinetic and potential energies, such as those in Eq.~\eqref{H}, for a system of $N$ particles. The quantization of the resulting Hamiltonian leads to
\be
\hat{H}_+'=\sum\limits_{j=1}^N\frac{\hat{p}_j'^2}{2m_j}+
\hat{V}(0,\hat{\mathbf{x}}_j')-\dot{x}_0\hat{P}',
\ee
where $\hat{P}'=\sum_j \hat{p}_j'$.
It follows that $m_j\,\ddot{\hat{x}}_j'=-\partial_{\hat{x}_j'}\hat{V}-m_j\,\ddot{x}_0$. The resemblance of these results with those expressed by Eq.~\eqref{Hrel} and \eqref{acceleration} is apparent. Yet, note that the Hamiltonian
\be
\hat{H}_-'=\sum\limits_{j=1}^N\frac{\hat{p}_j'^2}{2m_j}+\hat{V}(0,\hat{\mathbf{x}}_j')
+M\ddot{x}_0 \hat{X}',
\ee
with $\hat{X}'=\frac{1}{M}\sum_jm_j\hat{x}_j'$, yields the same equation of motion for the $j$-th particle. Particularly in this case we have an illustration of the {\em equivalence principle}, according to which an observer within a frame with external acceleration $\ddot{x}_0$ predicts that the center of mass of the system is immersed in a local gravitational field with energy $M\ddot{x}_0\hat{X}'$.

Although the equations of motion produced by the above Hamiltonians are identical, the interpretation in each case is slightly different. For $\hat{H}'_+$ the fictitious force $-m\ddot{x}_0$ appears because the velocity $\dot{\hat{x}}_j'=\hat{p}_j'/m_j-\dot{x}_0$ depends on a kind of vector potential. On the other hand, for $\hat{H}'_-$ the noninertial force comes from a fictitious gravitational potential. Actually, it is easy to check that $\hat{H}'_+$ and $\hat{H}'_-$ are specializations of the one-parameter Hamiltonian
\be
\hat{H}'_{\epsilon}=\sum\limits_{k=1}^N\frac{\hat{p}_k'^2}{2m_k}+\hat{V}(0,\hat{\mathbf{x}}_k')-\text{\scriptsize $\frac{(1+\epsilon)}{2}$}\dot{x}_0\hat{P}'+\text{\scriptsize $\frac{(1-\epsilon)}{2}$}M\ddot{x}_0 \hat{X}',
\ee
which produces $m_k\ddot{\hat{x}}_k'=-\partial_{\hat{x}_k'}\hat{V}-m_k\ddot{x}_0$ regardless the choice for $\epsilon$. These remarks show that as far as a change of classical frames is concerned the noninertial effects can be suitably accounted for via either vector potentials or fictitious gravitational fields, or even via both simultaneously. 

Naturally, one may wonder whether this gauge-like invariance applies to the physics relative to a quantum reference frame, as given by Eq.~\eqref{Hrel}. An inspection of the structure underlying $\hat{H}'_{\epsilon}$ shows that this cannot be the case for $\hat{H}_{rel}$. For instance, one could try, by analogy, to express the latter in terms of a potential energy $\sum_k m_k \hat{x}_{r_k}d\hat{\Pi}/dt$. However, since $i\hbar \,d\hat{\Pi}/dt=[\hat{\Pi},\hat{V}]=f(\hat{\mathbf{x}}_r)$, the resulting energy would be nonlinear in the position operators and the notion of a local gravitational field, as in $\sum_k m_k \hat{x}_k'\ddot{x}_0$, would not maintain. Moreover, it is easy to check that this attempt would not generate the correct equations of motion. Thus, the structure offered by the Hamiltonian~\eqref{Hrel}, with velocities given by Eq.~\eqref{velocity}, relying on the analogy with vector potentials, turns out to be the most natural canonical formulation for the physics within a noninertial quantum reference frame. Although not as intuitive as a description based on gravitational fields, it succeeds in producing the correct predictions for the dynamics of particles in accelerated frames. 

As a final remark, we note that besides accounting for the noninertial effects in the Heisenberg equations the kinetic term $\hat{\Pi}^2/2m_0$ is the ingredient responsible for the apparent nonlocality found in the vector state whenever the reference particle interacts with the system (see Sec.~\ref{rnonlocality}). This can be easily illustrated in a simple system, say with equal-mass particles 0, 1, and 2, where particle 0 interacts only with particle 1. Suppose that a potential $\hat{V}(\hat{x}_1-\hat{x}_0)=\hat{V}(\hat{x}_{r_1})$ is able to dynamically transform the state $\ket{\psi}_{t_0}=\ket{0}_0\ket{0}_1\ket{3a}_2$ into
\be
\ket{\psi}_t&=&\left(\frac{\ket{-a}_0\ket{a}_1+\ket{a}_0\ket{-a}_1}{\sqrt{2}}\right)\ket{3a}_2\nonumber \\ &=&\left(\frac{\ket{a,2a,4a}+\ket{a,-2a,2a}}{\sqrt{2}}\right)_{\cm, r_1, r_2}.
\ee
Clearly, while $\hat{V}(\hat{x}_{r_1})$ acts locally in the relative partitions the term $\hat{\Pi}^2$ can produce the above relative entanglement   through the coupling $\hat{\pi}_1\hat{\pi_2}$.

\section{Conclusion}\label{conclusion}

In this paper we aimed to carry on the program initiated in Refs.~\cite{everett57,aharonov155,aharonov158,aharonov73,aharonov74,aharonov84} and retaken in Ref.~\cite{angelo11} which aims to formulate a {\em relational} version of quantum mechanics. This includes to abandon the primitive notion of a classical reference frame in the background of the theory so as to obtain a self-contained quantum formulation free of any classical stuff. Here the goal was achieved for spatial degrees of freedom through a usual strategy, which consisted in starting from the ordinary formalism---presumably valid from the perspective of an absolute {\em inertial} frame of reference external to the system---and then to move to center of mass plus relative coordinates.

Many interesting aspects underlying the kinematics in quantum reference frames were unveiled and understood. First, we showed that the state seen by a quantum particle fundamentally depends on the masses of the particles, a distinct nonclassical effect. Second, we found out the characteristics that make the quantum reference frame equivalent to the classical absolute one, thus identifying how the notion of classicality emerges from the quantum substratum. Third, we discussed the emergent nonlocality which derives from the lightness of the quantum frame. As pointed out (see also Ref.~\cite{angelo11}), the Hilbert space associated with the relative partitions is fundamentally nonseparable. Fourth, we analyzed how the uncertainty relations and external correlations manifest in the perspective of the reference particle. Next, some puzzling questions were investigated which highlighted the need for correctly accounting for the measuring device as a quantum particle of the system. We showed that in theses circumstances the complementary principle is verified by all observers. Moreover, it follows from our approach that no detector is {\em absolutely classical}, as it is not spatially localized from all perspectives. Finally, it was shown throughout the presentation how to distinguish realistic correlations from those derived from the absolute motion of the quantum frame.

As far as the dynamics is concerned, we derived a Hamiltonian formulation for quantum reference frames, which are, in essence, {\em noninertial}. In reference to Aharonov's works~\cite{aharonov73,aharonov74,aharonov84}, we derived a canonical formulation of the physics seen from the perspective of noninertial quantum frames in which fictitious forces manifest formally as vector potentials. The resulting relative Hamiltonian, conveniently decoupled from the center of mass, exhibits momentum-momentum coupling which correctly accounts for the relational nonlocality.

In conclusion, we have shown how to suitably describe and interpret the physics of spatial degrees of freedom from the perspective of a noninertial quantum frame of reference. Although our arguments have been given for one-dimensional systems, we do not expect the extension to higher spatial dimensions to be much more involved technically. Conceptually, however, further puzzling issues are likely to appear. As far as we can see, the conjugation of this contribution to Refs.~\cite{aharonov84,angelo11} constitutes a rather complete conceptual framework for the physics seen from a quantum reference frame.

\section*{Acknowledgments}
The authors acknowledge financial support from INCT-IQ (CNPq/Brazil).


\appendix
\section{The Aharonov-Kaufherr transformation} 
\label{comparison_AK}

The canonical transformation proposed by Aharonov and Kaufherr in Ref.~\cite{aharonov84}, for a system of two particles, reads
\be\begin{array}{lll}
\hat{q}_0=\hat{x}_0, &\qquad & \hat{\pi}_0=\hat{p}_0+\hat{p}_1, \\
\hat{q}_1=\hat{x}_1-\hat{x}_0, & & \hat{\pi}_1=\hat{p}_1.
\end{array}
\label{AKtransf}
\ee
Notably, the new canonical momentum $\hat{\pi}_1$ keeps referring to the absolute frame. Given the change of basis
\be
\ket{x_0}_{x_0}\ket{x_1}_{x_1}=\ket{x_0}_{q_0}\ket{x_1-x_0}_{q_1},\nonumber
\ee
it is easy to show that a generic two-particle state 
\be
\ket{\psi}=\int dx_0dx_1~\psi(x_0,x_1)~\ket{x_0}_{x_0}\ket{x_1}_{x_1}
\label{psix0x1}
\ee
rewrites
\be
\ket{\psi}=\int dq_0dq_1~\Big[e^{q_0\partial_{q_1}}\psi(q_0,q_1)\Big]~\ket{q_0}_{q_0}\ket{q_1}_{q_1}.\nonumber
\ee
The elements of the reduced density matrix $\hat{\rho}_{q_1}=\text{Tr}_{0}\left(\ket{\psi}\bra{\psi}\right)$ are given by
\be
\bra{\chi}\hat{\rho}_{q_1}\ket{\chi+\delta}=\int du~e^{u\partial_{\chi}}\Big[ \psi(u,\chi)\psi^*(u,\chi+\delta)\Big],\nonumber
\ee
which is insensitive to the masses, in contrast with our previous result \eqref{coherences}. The reason for this apparent conflict can be understood as follows.

One of the main lessons of Ref.~\cite{angelo11} is that the Hilbert space of light frames is intrinsically nonlocal. The immediate consequence of this fact is that our usual representations of quantum states must be reviewed carefully, as illustrated throughout this paper. In particular, it is shown that descriptions given in terms of relative coordinates, without any assessment of the physical content of the correlations, may lead to all sort of nonphysical preconceptions. It follows that the safer description turns out to be the one accounting for information about the momenta as well. 

One option is to rewrite Eq.~\eqref{psix0x1} by inserting a momentum basis so as to get
\be
\ket{\psi}=\int \frac{d^2\vec{x}\, d^2\vec{p}}{2\pi\hbar}~e^{-i\vec{p}\cdot\vec{x}/\hbar}~\psi(\vec{x})~\ket{\vec{p}\,}_{p},\nonumber
\ee
where $d^2\vec{p}=dp_0dp_1$ and $\ket{\vec{p}}_{p}=\ket{p_0}_{p_0}\ket{p_1}_{p_1}$, with equivalent relations for $\vec{x}$. In terms of the new variables, as defined by Eq.~\eqref{AKtransf}, one gets
\be
\ket{\psi}=\int \frac{d^2\vec{q}\, d^2\vec{\pi}}{2\pi\hbar}~e^{-i\vec{\pi}\cdot\vec{q}/\hbar}~\psi(\vec{x}(\vec{q}))~\ket{\vec{\pi}\,}_{\pi}.\nonumber
\ee
The crucial point is that this representation does not refer solely to the relative physics. In fact, by Eq.~\eqref{AKtransf} one sees that $\pi_1$ actually is the absolute momentum. The purely relative description can be obtained by noting that $\pi_0=p_{\cm}$, $\pi_1=p_1=p_r+\frac{m_1}{M}p_{\cm}$, and $\ket{\pi_0}_{\pi_0}\ket{\pi_1}_{\pi_1}=\ket{\pi_0}_{p_{\cm}}\ket{\pi_1-\frac{m_1}{M}\pi_0}_{p_r}$. Using the notation $\vec{\wp}=(p_{\cm},p_r)$ one shows that 
\be
\ket{\psi}=\int \frac{d^2\vec{q}\, d^2\vec{\wp}}{2\pi\hbar}~e^{-i\vec{\wp}\cdot\vec{q}/\hbar}~e^{-\frac{i p_{\cm}q_1}{\hbar}\frac{m_1}{M}}~\psi(\vec{x}(\vec{q}))~\ket{\vec{\wp}\,}_{\wp},\nonumber
\ee
which explicitly depends on the masses. From this result one may infer that any representation involving relative coordinates and momenta will necessarily depend on the masses.


\begin{thebibliography}{99}
\bibitem{note2} Research on the measurement process is in progress. Further results and interpretations underlying the uncertainty relations and their links with the measurement will be published elsewhere.
\end{thebibliography}
\end{document}